\documentclass[%
	aps, prb, superscriptaddress, 
	reprint, 
	nolinenumbers,
	longbibliography, 
	citeautoscript, footinbib,
	]{revtex4-2}

\usepackage{graphicx}
\usepackage{siunitx}
	
\usepackage{booktabs}
\usepackage{multirow}
\usepackage{tabularx}
\usepackage{array}

\usepackage{wasysym}

\renewcommand*{\epsilon}{\varepsilon}
\renewcommand*{\vec}[1]{\mathbf{#1}}

\DeclareTextCommand{\permil}{T1}{\%\char24}
\DeclareUnicodeCharacter{00BE}{\threequarter}
	
\usepackage{xcolor}

\newcommand{\MnGeO}{Mn$_2$GeO$_4$}

\newcommand{\ignore}[1]{}

\usepackage[normalem]{ulem} 

\begin{document}

\title{Magnetoelectric initialisation of multiferroic \MnGeO}

\author{Na\"emi Leo$^{\ast}$}
\affiliation{Department of Materials, ETH Zurich, 8093 Zurich, Switzerland}
\affiliation{Department of Physics, School of Science, Loughborough University, LE11 3TU Loughborough, United Kingdom}
\email{n.leo@lboro.ac.uk}

\author{Jonathan S.~White}
\author{Michel Kenzelmann}
\affiliation{Laboratory for Neutron Scattering and Imaging, Paul Scherrer Institut, 5232 Villigen, Switzerland}

\author{Takashi Honda}
\affiliation{Institute of Materials Structure Science, High Energy Accelerator Research Organization (KEK), Tsukuba, Ibaraki 305-0801, Japan}

\author{Tsuyoshi Kimura}
\affiliation{Department of Applied Physics, University of Tokyo, Bunkyo-ku, Tokyo 113-8656, Japan}

\author{Dennis Meier}
\affiliation{Department of Materials Science and Engineering, Norwegian University of Science and Technology (NTNU), 7491 Trondheim, Norway}
\affiliation{Center for Quantum Spintronics, Department of Physics, Norwegian University of Science and Technology (NTNU), 7491 Trondheim, Norway}

\author{Manfred Fiebig}
\affiliation{Laboratory for Multifunctional Ferroic Materials, Department of Materials, ETH Zurich, 8093 Zurich, Switzerland}

\begin{abstract}

	Magnetoelectric multiferroics promise direct cross-control between coexisting ferroelectric and ferromagnetic orders, which is of interest for applications in magnetism and spintronics. 
	A particularly interesting type of cross-control is found in spin-spiral multiferroic \MnGeO, where a ferroelectric multi-domain distribution can be globally inverted by a single magnetic field sweep. 
	In this work we consider the initial domain evolution from zero-field cooling, imaging the evolution of domains under both magnetic and electric fields via optical second harmonic generation.
	We find that polarization and magnetization domains form independently when entering the multiferroic phase, and a single deterministic initialisation procedure, spanning three quarters of a field cycle, is required to achieve reliable magnetoelectric cross-coupling. 
	This initialisation behaviour originates from a deterministic pathway from metastable to equilibrium domain patterns, in contrast to more common and less reliable domain "training" procedures that require repeated field cycles. 
	Understanding the initial domain evolution thus enables reliable cross-control in magnetoelectric devices with highly interlinked order parameters.

\end{abstract}


\maketitle

Magnetoelectric engineering, such as electric-field-induced switching of a magnetic order, underlies much of the recent research on multiferroics, that is, on materials with coexisting magnetic and ferroelectric orders within the same phase \cite{1994Schmid,2005Spaldin,2010Spaldin,2016Fiebig,2019Spaldin}.
Coexisting multiple order parameters can facilitate pronounced magnetoelectric cross-coupling, and in particular rigid one-to-one coupling between macroscopic magnetisation and electric polarisation is highly desirable for functional devices \cite{2005Fiebig,2006Eerenstein}.

Especially strong magnetoelectric effects are observed in antiferromagnetic spin-spiral multiferroics, in which the inversion-symmetry-breaking compensated spin order is directly responsible for the emergence of a spontaneous electric polarization \cite{2007Cheong,2009Wang,2011Arima}.
A particular case of such magnetoelectric domain cross-control has been demonstrated in conical-spiral multiferroic \MnGeO\ \cite{2012White,2012Honda,2014Honda,2017Honda}. As we showed in a previous work \cite{2018Leo_trilinear}, a trilinear magnetoelectric coupling of order-parameter terms enables the complete inversion of a ferroelectric multi-domain pattern by a homogeneous magnetic field, where the the spontaneous polarization is reversed in every position on the sample, while the domain pattern as such remains unchanged. 
This reliable interconversion is, however, observed for a fully equilibriated domain configuration only.

Complicating the picture field-driven coupling, however, are often-observed irreversible changes of the domain pattern upon the initial field cycles after zero-field cooling into the ordered state.
This behaviour is known as ``training" or ``annealing" procedures \cite{2013Zhou, 2018DeLuca, 2021Muller, 2023Saini} and are little studied in multiferroics, limiting the practical value of their magnetoelectric coupling.
%

In this work, we investigate the initial field-driven domain evolution in \MnGeO\ after zero-field cooling, using optical second harmonic generation (SHG) as spatially resolving domain-imaging technique. 
We find that the initial ferroelectric and ferromagnetic domains form independently, and that an irreversible initialisation procedure is required to enable the repeatable magnetoelectric cross-coupling. 
The initialisation procedure described here differs markedly from that of other ``training" procedures in that is is fully deterministic and requires a single field cycle only.
We explain the domain evolution on a phenomenological level of global free-energy minimisation of a trilinear order-parameter coupling term, and propose a microscopic mechanism.
Our findings emphasize the importance of a comprehensive understanding of the domain evolution away from equilibrium for obtaining fully reproducible magnetoelectric control, a prerequisite for reliable multiferroic devices.

\section{Ferroic Properties of M\symbol{110}$_2$G\symbol{101}O$_4$}
\label{sec:material_properties}

\begin{figure}
	\centering
	\includegraphics[width=86mm]{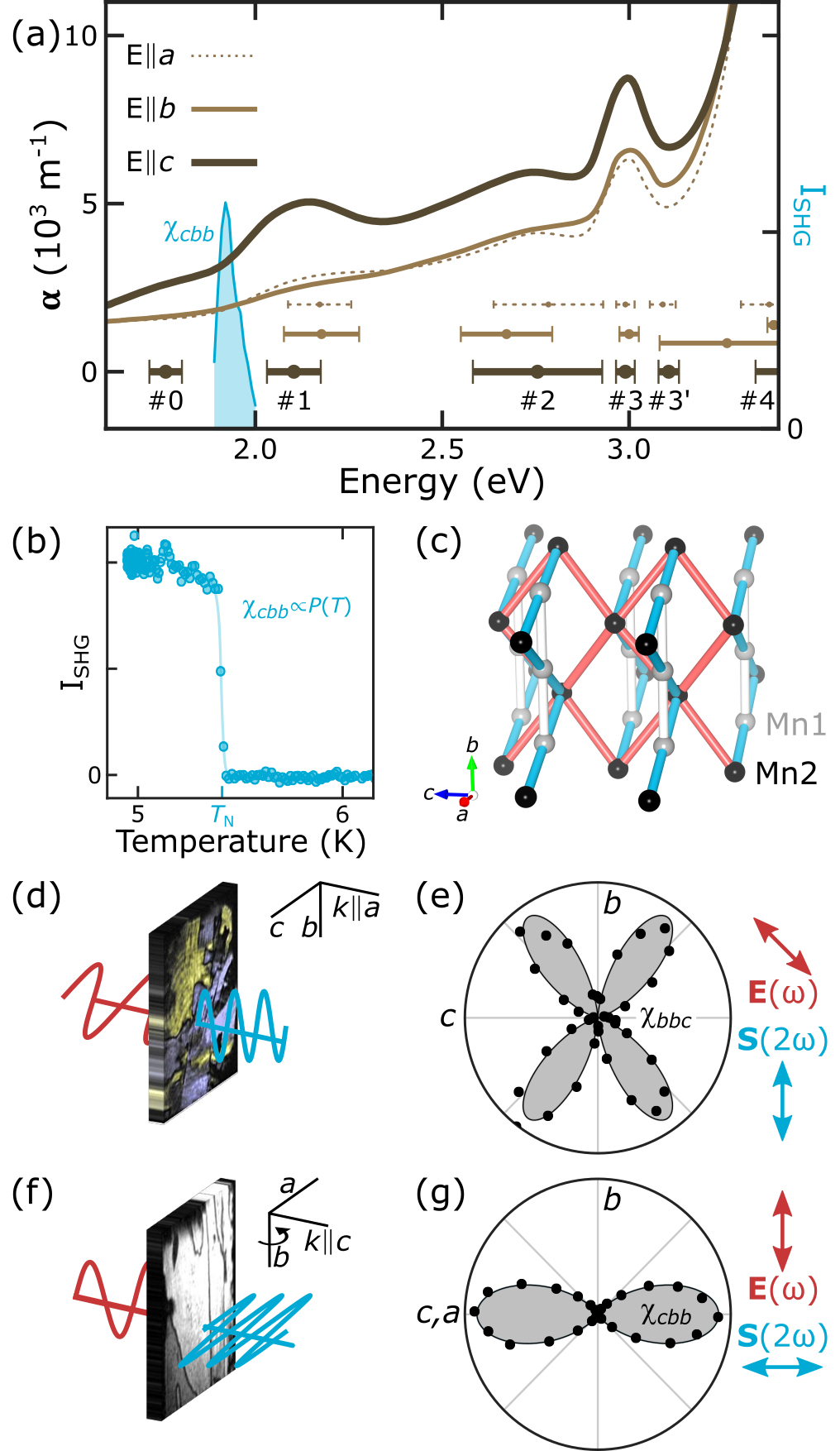} 
	\caption{%
		\textbf{Optical properties of \MnGeO.}
		(a)~Linear optical absorption coefficient $\alpha(\hbar\omega)$ with light polarized along the $a$, $b$, and $c$ axes of orthorhombic \MnGeO\ (linear scale on left). Central energies and width of the optical excitations are indicated by bars at the bottom (Tab.~\ref{tab:Mn2GeO4:optical_transitions}). \textbf{M1}
		The blue peak indicates the spectral dependence of the SHG response of $\chi_{cbb}$ at $2\hbar\omega$ (linear scale on right).
		(b)~Temperature dependence of the SHG signal $\chi_{cbb}\propto P$ at \SI{1.85}{eV}. The signal is present in the multiferroic phase below $T_{\rm N}$ only, and couples to the ferroelectric polarisation $P$.
		(c)~Crystal structure of \MnGeO, with centric Mn1 and acentric Mn2 sites in grey and black, respectively.
		(d,f)~SHG transmission setup for domain imaging of $a$-cut and $c$-cut samples, respectively. To observe the SHG response of the $c$-cut sample, it needs to be slightly rotated around the vertical $b$ axis so that $c$-polarized contributions to the light waves can be excited.
		(e,g)~Rotational SHG anisotropies, measured between crossed polarizers, of the (e)~$a$-cut and (g)~$c$-cut sample. \textbf{M2}
		For SHG components $\chi_{cbb}$ and $\chi_{bbc}$ the incident light field $\vec{E}(\omega)$ is polarized parallel to the Mn2-O-Mn2 bonds (red) that promote strong antiferromagnetic exchange. 
	}
	\label{fig:optics}
\end{figure}

The olivine compound \MnGeO\ crystallizes in the orthorhombic space group \textit{Pnma} with principal axes $a>b>c$ \cite{1970Creer}.
Its optical and magnetic properties are closely linked to the electronic configuration of the Mn$^{2+}$ ions \cite{Burns93}, which occupy two crystallographic sites with octahedral coordination. The Mn1 ions occupy centrosymmetric $4a$ positions, while the Mn2 ions are situated in acentric $4c$ positions, but with local mirror symmetry \cite{1970Creer}. 

\MnGeO\ exhibits a series of magnetic phase transitions \cite{2012White,2012Honda,2014Honda,2017Honda}. In the current work, we focus on the conical spin-spiral phase, which is entered through a first-order phase transition at about $T_{\rm N}=\SI{5.5}{K}$ \cite{2012Honda, 2012White,2013Volkov}.
The conical spin-spiral order in this low-temperature multiferroic phase is characterized by the coexistence of a multitude of order parameters associated with the commensurate wave vector $\vec{Q}_{\rm c}=0$ and the incommensurate wave vectors $\vec{Q}_{\rm A}=(0.136, 0.211, 0)$ and $\vec{Q}_{\rm B}=(0.136, -0.211, 0)$ \cite{2012Honda,2012White}.
Associated with the commensurate order parameterized by $\vec{Q}_{\rm c}$ are the antiferromagnetic order parameters $X_1$ and $X_3$. Here, $X_1\propto\mathcal{C}$ is purely antiferromagnetic, while $X_3$ is related to a small canted magnetization $M\parallel c$ with magnitude $M=\SI{7e-3}{\mu_B/Mn^{2+}}$. 
The incommensurate modulation associated with $\vec{Q}_{\rm ic}$ describes a cycloidally modulated spin structure and induces a spontaneous ``improper'' electric polarization $P\parallel M \parallel c$ with magnitude $P=\SI{16}{\micro C/m^2}$. 
The macroscopic magnetization $M$ and polarization $P$ can be completely switched to a ferromagnetic or ferroelectric single-domain state by magnetic fields  ($H_\mathrm{coerc}\approx\SI{100}{mT}\parallel c$) or electric fields \mbox{($E_\mathrm{coerc}\approx\SI{5}{MV/m}\parallel c$)}, respectively \cite{2012White,2017Honda}. 

In the multiferroic phase below $T_{\rm N}$, the spontaneous magnetization and polarization are strongly coupled. This has been demonstrated by the synchronous reversal of both magnetization and polarization in a magnetic field for a single-$P$--single-$M$ domain configuration \cite{2012White,2017Honda} as well as by the magnetic-field-driven inversion of a multi-$P$--single-$M$ domain pattern. In the latter case, the spatial distribution of the ferroelectric domains is retained while the local polarization is flipped \cite{2018Leo_trilinear}.
Macroscopically, the domain inversion has been attributed to a trilinear magnetoelectric coupling term \cite{2017Harris, 2018Leo_trilinear}. Microscopically, the simultaneous switching of polarization and magnetization in \MnGeO\ originates from a magnetic-field-induced directional flop of the Mn spin-spiral cone axis, which reverses both the net magnetization and the projected spiral helicity, and thus the spin-induced polarization \cite{2017Honda}.

\section{Methods}
\label{sec:methods}

Single crystals of \MnGeO\ were grown by the floating-zone method \cite{2010Tokunaga}, oriented by Laue diffraction, and cut into platelets with faces perpendicular to the crystallographic $a$ or $c$ axes. The samples were lapped and polished from both sides with silica slurry (Eminess Ultrasol 555). The final thickness was approximately \SI{70}{\micro m} ($a$-cut) and \SI{60}{\micro m} ($c$-cut).

Linear optical absorption spectra were measured at room temperature, using a microspectrometer (JASCO MSV-370). 
%
SHG experiments were performed with amplified \SI{120}{fs} laser pulses of variable photon energy in a transmission geometry, using a setup described elsewhere \cite{Fiebig05b}.
%
The spatial distribution of the ferroelectric domains was imaged with a tele\-photography lens (Soligor 135~mm, \mbox{1:3.5}) using a liquid-nitrogen-cooled CCD camera as the detector (Photometrics CH270). All imaging experiments were performed in the multiferroic phase at $T\approx\SI{4.5}{K}$, with the low-temperature sample environment provided by a liquid-helium-operated magnetic-field cryostat (Oxford Spectromag 4000).

\begin{figure}[tb] 
	\centering
	\includegraphics[width=86mm]{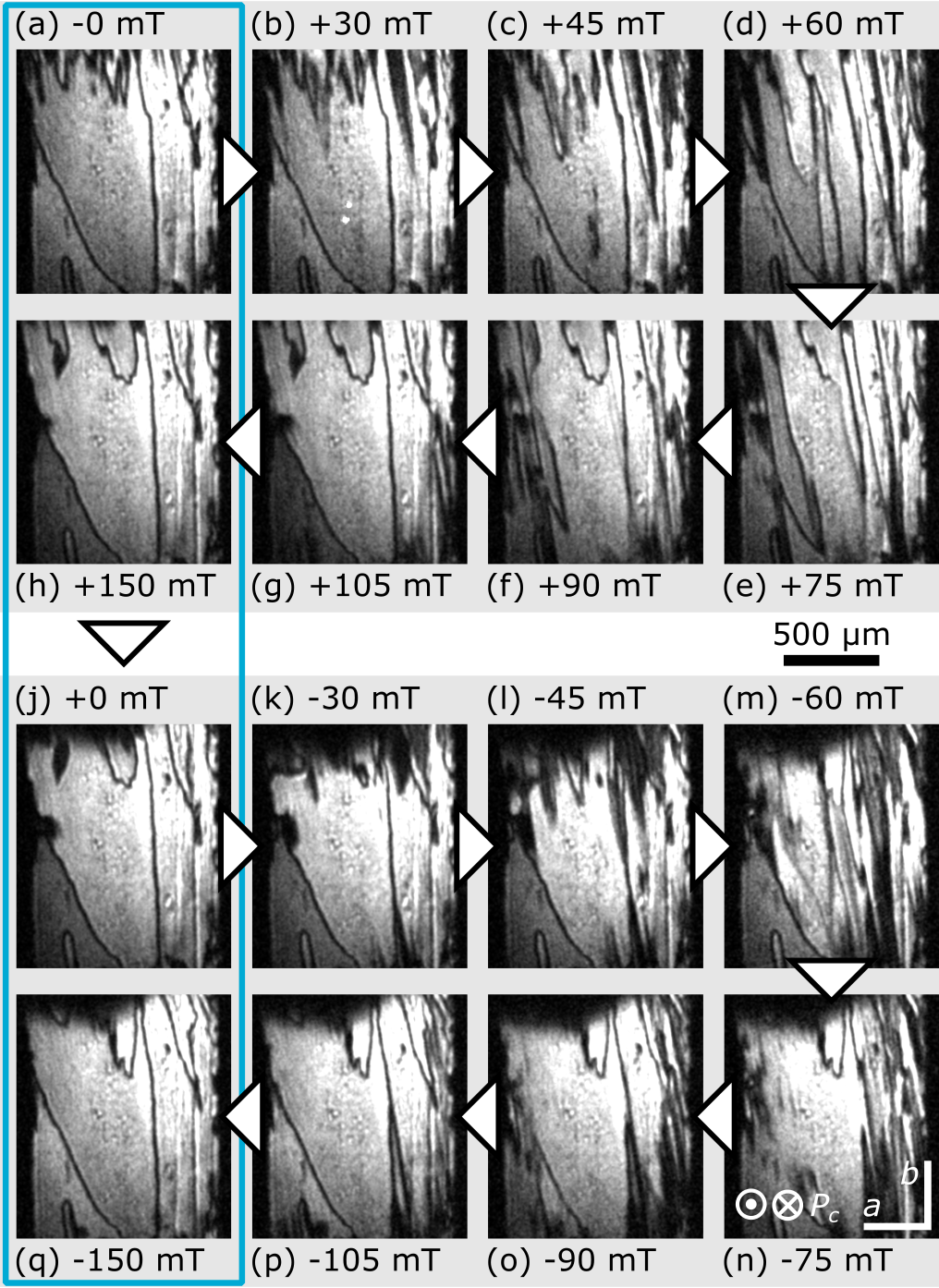}
	\caption{%
		\textbf{Polarization domains upon magnetization reversal.}
		Evolution of a multi-$P$--single-$M$ domain pattern obtained after an initial field cycling with (a-h)~rising and (j-q)~falling applied magnetic field $H_c$.
		Neighboring $\pm P$ domains appear with equal brightness separated by black lines. 
		Intermediate magnetic fields lead to multi-$P$--multi-$M$ domain patterns that differ in subsequent field cycles, as demonstrated by the transient ferroelectric domains. 
		After a complete magnetization reversal, that is, comparing panel (a) to (h) and (j) to (q), the domain pattern is largely recovered (first column), but with the polarity of each local domain inverted, i.e., $\pm P\rightarrow\mp P$.
		}
	\label{fig:magnetic-switching}
\end{figure}

To switch the spontaneous magnetization $M$ of \MnGeO, magnetic fields $H$ were applied along the crystallographic $c$ direction. To be able to observe the SHG signal of the $c$-cut platelet, the sample had to be rotated by approximately \SI{20}{\degree} around the vertical $b$ direction. 
This rotation leads to a perspective distortion of the SHG images. This is corrected by stretching the images in Fig.~\ref{fig:magnetic-switching} and Fig.~\ref{fig:electric-field-switching} along their horizontal direction.

To obtain electric-field poling within the multiferroic phase in the $c$-cut sample, transparent indium tin oxide (ITO) electrodes were sputtered onto both sides of the polished crystal, which allows application of electric fields of up to \SI{15}{MV/m} before electrical breakthroughs occur.
The ITO layer does not completely extend over each sample surface; hence the electric field $e\parallel c$ is applied in the overlap region only (highlighted blue in Fig.~\ref{fig:electric-field-switching}).

\section{Optical characterization / SHG and ferroelectric order}
\label{sec:optics}

\begin{table}[t] 
	\caption{%
		\textbf{Optical transitions, as observed in Fig.~\ref{fig:optics}(a).}
		The metal-oxygen charge-transfer edge CT, the low-energy excitation \#0, and crystal-field transitions \#1 to \#4 were fitted with Gaussian functions centred at $E_{\rm c}$ with amplitude $A$ and full width at half maximum $w$. A constant background with amplitude $A^\prime_0$ was also assumed.
	}
	\label{tab:Mn2GeO4:optical_transitions}
	\begin{center}
		\begin{tabularx}{0.475\textwidth}{l l @{\hskip 1em} l @{\hskip 1em} rrr @{\hskip 1em} p{0.2\textwidth}}
			\hline\hline
			&			&		
			& \multicolumn{1}{c}{$E\parallel a$} 
			& \multicolumn{1}{c}{$E\parallel b$} 
			& \multicolumn{1}{c}{$E\parallel c$} 
			& \\
			\hline\hline
			& $A_0^\prime$			& \si{m^{-1}} 	& 1393		& 1146		& 2310		& background \\	
			\hline \hline
			\multirow{3}{*}{\#0}	
			&	$E_{\rm c}$ 				& \si{eV}		& -- 		& -- 		& 1.760		& 	
			\multirow{3}{\linewidth}{ ~ } 
			\\
			&	$w$ 				& \si{eV}		& -- 		& -- 		& 0.205		& 	\\
			& 	$A$ 				& \si{m^{-1}}	& -- 		& -- 		& 271		& 	\\
			\hline
			\multirow{3}{*}{\#1}	
			&	$E_{\rm c}$ 				& \si{eV}		& 2.172		& 2.177		& 2.103		& 	
			\multirow{3}{\linewidth}{${}^6A_{1g}\rightarrow {}^4T_{1g}$}
			\\
			&	$w$ 				& \si{eV}		& 0.398		& 0.474 	& 0.339		& 	\\
			& 	$A$ 				& \si{m^{-1}}	& 1090		& 1017		& 2025		& 	\\
			\hline
			\multirow{3}{*}{\#2}	
			&	$E_{\rm c}$ 				& \si{eV}		& 2.784 	& 2.672		& 2.755		& 
			\multirow{3}{\linewidth}{${}^6A_{1g}\rightarrow {}^4T_{2g}$}
			\\
			&	$w$ 				& \si{eV}		& 0.690 	& 0.576 	& 0.815		& 	\\
			& 	$A$ 				& \si{m^{-1}}	& 4360		& 2406		& 7955		& 	\\
			\hline
			\multirow{3}{*}{\#3}	
			&	$E_{\rm c}$ 				& \si{eV}		& 2.990 	& 3.000 	& 2.990		& 	
			\multirow{3}{\linewidth}{${}^6A_{1g}\newline{ }\,\rightarrow {}^4E_g, {}^4A_{1g}$}
			\\
			&	$w$ 				& \si{eV}		& 0.116		& 0.122		& 0.118		& 	\\
			& 	$A$ 				& \si{m^{-1}}	& 681		& 621		& 948		& 	\\
			\hline
			\multirow{3}{*}{\#3$^\prime$}
			&	$E_{\rm c}$ 				& \si{eV}		& 3.090		& --		& 3.106		& 
			\multirow{3}{\linewidth}{(${}^6A_{1g}\rightarrow {}^4E_g$)}
			\\
			&	$w$ 				& \si{eV}		& 0.163		& --		& 0.129		& 	\\
			& 	$A$ 				&  \si{m^{-1}}	& 264		& --		& 201		& 	\\
			\hline
			\multirow{3}{*}{\#4$^\prime$}
			&	$E_{\rm c}$ 				& \si{eV}		& --		& 3.262		& --		& 	\\
			&	$w$ 				& \si{eV}		& --		&(0.850)	& --		&	\\
			& 	$A$ 				& \si{m^{-1}}	& --		& (6160)	& --		& 	\\
			\hline
			\multirow{3}{*}{\#4}
			&	$E_{\rm c}$ 				& \si{eV}		& 3.375		& 3.387		& 3.419		& 
			\multirow{3}{\linewidth}{${}^6A_{1g}\rightarrow {}^4T_{2g}$}
			\\
			&	$w$ 				& \si{eV}		& 0.357 	&(0.080)	& 0.380		& 	\\
			& 	$A$ 				& \si{m^{-1}}	& 1843		& (101)		& 2159		& 	\\
			\hline\hline
			\multirow{4}{*}{CT}
			&	$E_{\rm c}$ 				& \si{eV}		& 3.716		& 3.638		& 3.861		& 
			\multirow{4}{\linewidth}{charge transfer}
			\\
			&	$w$ 				& \si{eV}		& 0.482		& 0.432 	& 0.626		& 	\\
			& 	$A$ 				& \si{m^{-1}}	& 77364 	& 49271 	& 102906	&	\\
			&   $A_0$ 				& \si{m^{-1}}  	& 8821		& 10623 	& 11069		& 	\\
			\hline\hline
		\end{tabularx}
	\end{center}
\end{table}

The $3d^5$ electronic configuration of the Mn$^{2+}$ ions dominates the optical response in \MnGeO\, for which $d-d$ transitions are parity-forbidden \cite{1954Tanabe_b}. \MnGeO\ crystals appear reddish-brown, however, due to the low-energy tail of the lowest-lying Mn$\rightarrow$O charge-transfer excitation peaking around \SI{3.7}{eV}. 
The polarization-dependent linear spectra in Fig.~\ref{fig:optics}(a) show the strongest absorption for light polarized along the crystallographic $c$ direction, leading to a pronounced dichroism.
Each spectrum was fitted with a multi-Gauss function to extract the central energies $E_{\rm c}$ and the full-width-half-maximum peak widths $w$ of the crystal-field transitions \#1 to \#4. The transitions are indicated in Fig.~\ref{fig:optics}(a) and are listed in Tab.~\ref{tab:Mn2GeO4:optical_transitions} \cite{1954Tanabe_b,1970Manning,Burns93}.
The peak labeled \#0 at \SI{1.76}{eV} was found in the $c$-polarized spectrum only and is not related to a crystal-field transition; its origin is discussed below.

To spatially resolve domains and to image the effect of field-driven ferroic switching and magnetoelectric coupling in \MnGeO, we employ optical SHG, which describes the emission of light at frequency $2\omega$ from a material irradiated with laser light at frequency $\omega$, as described by the equation
\begin{equation}
\mathbf{S}(2\omega)\propto\hat{\chi}^{(2)}\mathbf{E}(\omega)\mathbf{E}(\omega)\, .
\label{eq:SHG}
\end{equation}
Here, the electric-field components of the incident light field, $\mathbf{E}(\omega)$, induce a source term $\mathbf{S}(2\omega)$ for the emitted frequency-doubled light wave. The tensor components of the non-linear susceptibility $\chi^{(2)}_{ijk}(2\omega)$ depend on the point-group symmetry of the host material and vanish in the leading electric-dipole order for materials with inversion symmetry \cite{Pershan63,Birss66}.
Therefore, SHG is especially suitable for investigating ferroelectrics and spin-spiral multiferroics \cite{Denev11,2009Meier}, where spontaneous polarization arises as an inversion-symmetry-breaking order parameter.

\ignore{In \MnGeO, a background-free SHG response is induced either by the presence of the spontaneous polarization $P$ in the multiferroic phase, or by application of an external electric field $E\|c$, both of which break the inversion symmetry of the system.}

In agreement with symmetry selection rules \cite{Birss66}, in $a$-cut \MnGeO\ samples we observe SHG signals arising from the component $\chi_{bbc}$ [Fig.~\ref{fig:optics}(e)] and $\chi_{cbb}$ [Fig.~\ref{fig:optics}(g)].
These signals appear with a jump below $T_{\rm N}$, shown for $\chi_{cbb}$ in Fig.~\ref{fig:optics}(b), showing a similar temperature dependence as the spontaneous polarization $P$ \cite{2012White}.
In addition, we find an electric-field-induced second-harmonic (EFISH) contribution with an amplitude proportional to the strength of the applied external field $E\|c$ and independent of the sample temperature.

The spontaneous SHG signal, blue peak in Fig.~\ref{fig:optics}(a), and the EFISH contribution with $E\|c$ exhibit the same spectral dependence, peaking for photon energies $2\hbar\omega$ around \SI{1.90}{eV}.

Interestingly, for both observed SHG components, the polarization of the incoming light field  $\vec{E}(\omega)$ is parallel to the Mn2-O-Mn2 bonds promoting strong antiferromagnetic exchange and linking neighboring zigzag chains [red in Fig.~\ref{fig:optics}(c)] \cite{2008Bulaevskii, 2012Honda, 2012White, 2013Volkov}.
This particular polarization dependence and the spectral vicinity of the SHG peak to the absorption peak \#0 indicate that both linear and nonlinear responses could be linked to pairwise spin-flip transitions enhanced by exchange interactions between neighboring Mn$^{2+}$ ions \cite{1972Ferguson}.

The SHG signals present in the multiferroic phase of \MnGeO, which are sensitive to the spontaneous polarization, can be used for the background-free observation of the ferroelectric multi-domain pattern with a spatial resolution down to about \SI{1}{\micro m}. 
Neighboring $\pm P$ domains typically appear in SHG images with constant brightness $I_\text{SHG}\propto|\chi_\text{SHG}|^2\propto|\pm P|^2$, separated by dark lines that mark the position of the ferroelectric domain walls. These lines are caused by local destructive interference between SHG contributions from neighboring domains due to the $180^{\circ}$ phase shift ($\leftrightarrow\pm P$) associated with these.
For the $a$-cut sample, interference with a reference EFISH signal was used to generate a domain contrast (to distinguish the ferroelectric signal from crystallographic defects), with $\pm P$ domains highlighted in yellow and blue, respectively.

\begin{figure}[tb] 
	\begin{center}
		\includegraphics[width=86mm]{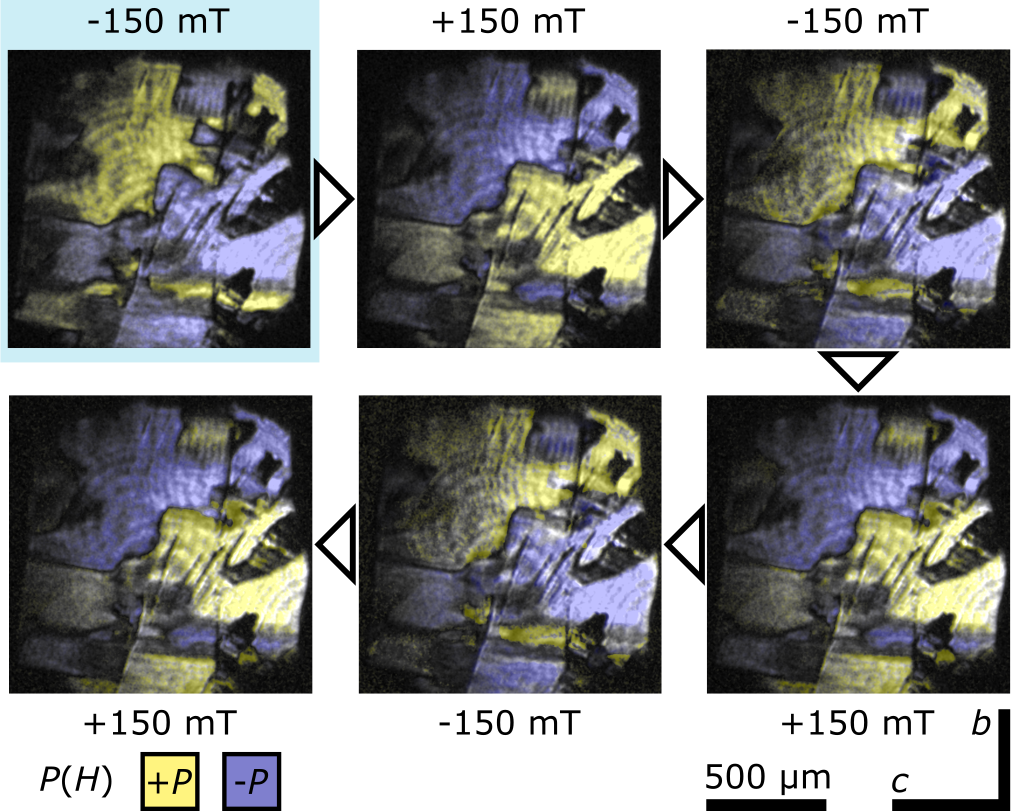}
	\end{center}
		\caption{%
			\textbf{Repeatability of magnetoelectric domain inversion.}
			Five subsequent reversals of the ferroelectric domain pattern, indicated by the interchange of blue and yellow contrast that indicate domains of opposite polarity $\pm P\parallel c$, are shown for switched magnetization $M(H_c)$.
			The initial domain configuration, highlighted in light blue, corresponds to the configuration shown in Fig.~\ref{fig:contrast-reversal}(d), obtained after a magneto\-electric initialization procedure from a zero-field-cooled domain configuration.
			}
		\label{fig:contrast-reversal-supplement}
\end{figure}

\begin{figure*}[tb] 
	\begin{center}
		\includegraphics[width=178mm]{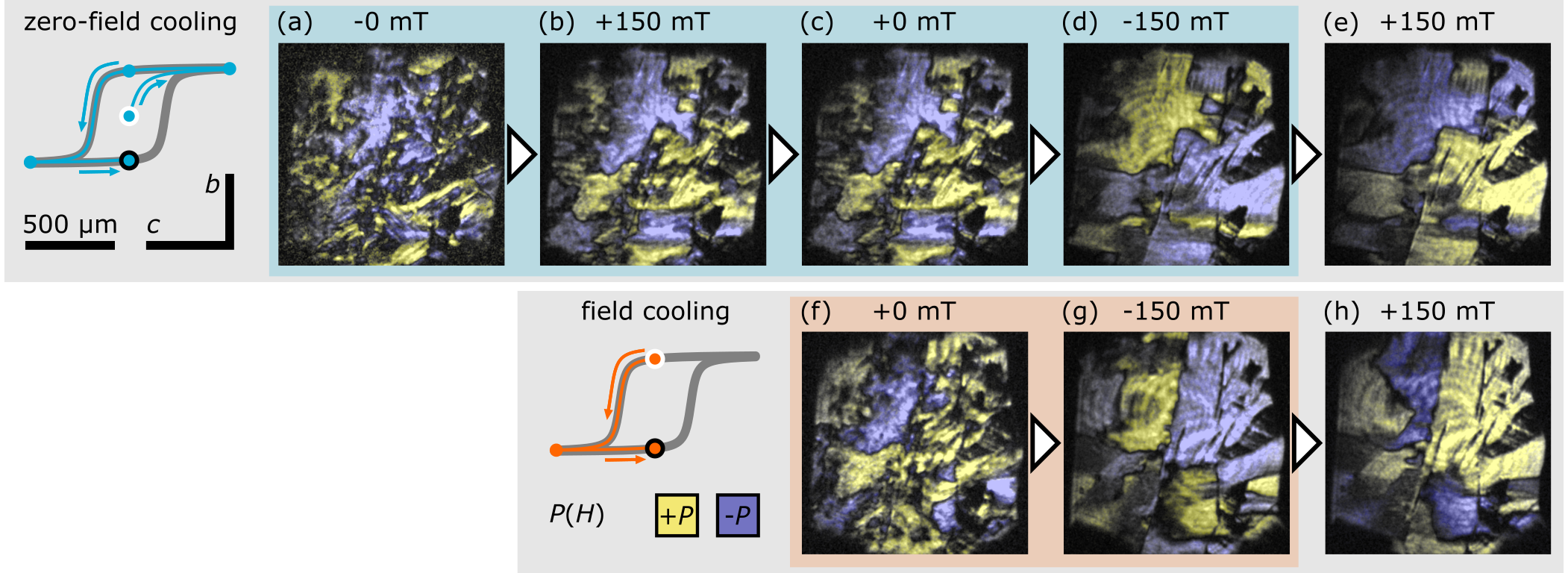}
		\caption{%
			\textbf{Initial domain evolution.}
			Ferroelectric domains in $a$-cut \MnGeO\ after (a-e)~zero-field cooling and (f-h)~$E_c$-$H_c$-field cooling at \SI{0.6}{MV/m} and \SI{500}{mT}. Blue and yellow areas denote domains of opposite polarity $\pm P$.
			Both morphology and size of domains change upon the first field-poling steps, before repeatable magnetoelectric domain inversion is achieved. This indicates an initial non-equilibrium distribution of $P$, $M$, and antiferromagnetic $\mathcal{C}$ domains.
			(f-h)~Similar changes upon the first magnetization reversal are observed after field-cooling in intermediate fields (highlighted in orange).
			(a-e)~After zero-field cooling (state with white outline), intermittent domain patterns changing in morphology and typical length scales are observed. This indicates an initial non-equilibrium distribution of $P$, $M$ and antiferromagnetic $\mathcal{C}$ domains.
			After a full field cycle (blue path), a remanent domain state that shows fully reversible magnetoelectric inter-conversion is reached (state with black outline). 
			(f-h)~After field cooling, an initial magnetic field induces similar domain changes as in (c-e) before stabilization of the magnetoelectric inter-conversion; in this case, only half a magnetic field cycle is required (orange path).
			Minor variations in (g,h) can be attributed to temperature fluctuations close to $T_{\rm N}$, possibly in combination with strain.
			}
		\label{fig:contrast-reversal}
	\end{center}
\end{figure*}

\section{Magnetoelectric domain inter-conversion}
\label{sec:imaging}

As shown in previous work \cite{2012White,2017Honda}, repeated application of a magnetic field along $c$ not only switches the magnetization of \MnGeO\ but simultaneously reverses the improper polarization. This effect leads to inversion of the ferroelectric domain pattern with one integral magnetic field sweep. Locally, each polarization state is inverted, $\pm P\rightarrow\mp P$, while the position of the ferroelectric domain walls as such is left unchanged \cite{2018Leo_trilinear}.

The associated magnetoelectric domain interconversion, which results in the transition of a multi-($\mp P$)--single-($-M$) domain pattern to its inverted multi-($\pm P$)--single-($+M$) state, is demonstrated for a $c$-cut \MnGeO\ sample in Figs.~\ref{fig:magnetic-switching}(a,h) and \ref{fig:magnetic-switching}(j,q).
Note how the final domain pattern in panel~(q) still resembles that of the initial pattern in panel~(a). Minor deviations occur only towards the edges of the image where the sample is mounted, which points to the disruptive influence of sample strain. The magnetoelectric inversion of the domain pattern is highly repeatable, as shown for five subsequent magnetic field reversals in Fig.~\ref{fig:contrast-reversal-supplement}.

Regardless of the repeatable interconversion, in intermediate fields [Figs.~\ref{fig:magnetic-switching}(b-f,k-p)], additional needle-like domains appear at the top of the sample and grow downward. 
As the spatial distribution and temporal evolution of domains in this transient multi-($P$)--multi-($M$) configuration is different in each poling cycle, the recovery of the initial domain state cannot be attributed to local pinning centers or other memory effects, and instead is governed by the intrinsic global magnetoelectric coupling.

To discuss the underlying cross-coupling effects, two switching mechanisms need to be distinguished in \MnGeO. These are represented by bilinear and trilinear coupling terms as contributions to the free energy, which must be minimized in equilibrium. 
First, the bilinear coupling terms describe how the applied fields act directly on the conjugated order parameter, that is, magnetization $M(H)$ and polarization $P(E)$, via 
\begin{equation}
	F_\mathrm{M} \propto -M(H) \cdot H < 0 
	\quad \mathrm{and} \quad
	F_\mathrm{E} \propto -P(E) \cdot E < 0 
	\, .
	\label{eq:bilinear}
\end{equation}
Second, the magnetoelectric trilinear coupling term is associated with the response of non-conjugated macroscopic order parameters according to \cite{2017Honda, 2017Harris,2018Leo_trilinear}:
\begin{equation}
F_{\text{ME}} \propto -\mathcal{C} \cdot M(H) \cdot P(E) < 0 \, .
\label{eq:trilinear}
\end{equation}
Due to the compensated spin order, the antiferromagnetic domains associated with the parameter $\mathcal{C}$ (a combination of commensurate and incommensurate order parameters, see \cite{2017Honda,2018Leo_trilinear}) are robust against moderate electric and magnetic fields. As a consequence of this stability of $\mathcal{C}$, when $M$ is switched through $H$ via the bilinear contribution $F_\mathrm{M}$ in Eq.~(\ref{eq:bilinear}), $P$ needs to reverse simultaneously in order to satisfy minimization of the trilinear free-energy contribution $F_{\text{ME}}$ in Eq.~(\ref{eq:trilinear}). Conversely, if $P$ is switched by an electric field $E$, the magnetization $M$ must reverse at the same time.

\subsection{Initialisation of the magnetoelectric inversion}
\label{sec:imaging:annealing}

In contrast to the highly-repeatable domain inter-conversion, the initial evolution of a of a multi-$P$--multi-$M$ domain pattern after zero-field cooling is strikingly different.
Specifically, we find that upon zero-field cooling to the multiferroic phase ($H=E=0$), polarization ($P$) and magnetization ($M$) domains form independently.

As shown in Fig.~\ref{fig:contrast-reversal}(a), the initial multi-$P$--multi-$M$ domain pattern features small $\pm P$ domains with predominantly diagonally oriented boundaries. An applied magnetic field converts the initial state to a multi-$P$--single-$M$ domain pattern. Due to the magnetoelectric coupling parameterized by Eq.~(\ref{eq:trilinear}), the ferroelectric domain pattern is expected to change in the regions where the magnetization has been switched. Indeed, as shown in Fig.~\ref{fig:contrast-reversal}(b,c), after $H$-field poling, the $\pm P$ domains appear larger, with boundaries preferably aligned with the horizontal (polar) $c$ axis, and slightly inclined to the vertical $b$ direction. 
The magnetic-field-driven change in the morphology and size of $P$ domain pattern hence confirms that the formation of independent distribution of ferroelectric and ferromagnetic multi-domain patterns upon entering the multiferroic phase.

To obtain the repeatable global magnetoelectric domain inter-conversion discussed above a \textit{deterministic} multi-step magnetic-field initialization procedure is required.
For its description in the following we will distinguish the effects of \textit{domain poling}, that is, the transformation of the virgin multi-domain pattern obtained after zero-field cooling into a single-domain configuration, and \textit{domain switching}, describing the inversion of a single-domain pattern via an intermittent multi-domain state as described in Fig.~\ref{fig:magnetic-switching}.
In other words, \textit{poling} affects only a subset of domains, whereas \textit{switching} inverts \textit{all} domains.

The image sequence shown in Fig.~\ref{fig:contrast-reversal}(a-c) represents the initial domain poling toward a single-($M$) domain configuration.
Upon the first global switching of the magnetization $+M\rightarrow -M$, Fig.~\ref{fig:contrast-reversal}(c,d), the majority of $\pm P$ domains invert to $\mp P$; however, this step is still distinct from the rigid and repeatable inter-conversion discussed above. In particular, some domains (especially smaller ones) disappear and the domain walls straighten considerably. Consequently, the resulting multi-($\mp P$)--single-($-M$) domain pattern is different from the previous multi-($\pm P$)--single-($+M$) domain pattern, in contrast to what would naïvely be expected from the magneto\-electric coupling term in Eq.~(\ref{eq:trilinear}).

A similar initialization behaviour is observed even if the sample has been initially cooled in a moderate magnetic field of $H=\SI{500}{mT}$, as shown in Fig.~\ref{fig:contrast-reversal}(f,g). Only subsequent reversals reveal the magnetic-field-driven equilibrium inversion of the ferroelectric domain pattern, Fig.~\ref{fig:contrast-reversal}(d,e) and Fig.~\ref{fig:contrast-reversal}(g,h).

The observed changes indicate that the relationship between $P$, $M$, and $\mathcal{C}$ domain populations is not yet in equilibrium after cooling in zero field or with a moderate bias field applied, as illustrated by the blue and orange paths on the hysteresis loops in Fig.~\ref{fig:contrast-reversal}. Instead, obtaining the equilibrium state of field-induced domain inversion requires one full, respectively one half, magnetic-field initialization cycle.
After these initial steps, the system globally minimizes the magnetoelectric coupling term in Eq.~(\ref{eq:trilinear}) and allows for repeated domain inversion in successive switching cycles, as illustrated in Fig.~\ref{fig:contrast-reversal-supplement}.

\subsection{Electric-field control of multiferroic domains}
\label{sec:imaging:efields}

\begin{figure}[tb] 
	\centering
	\includegraphics[width=86mm]{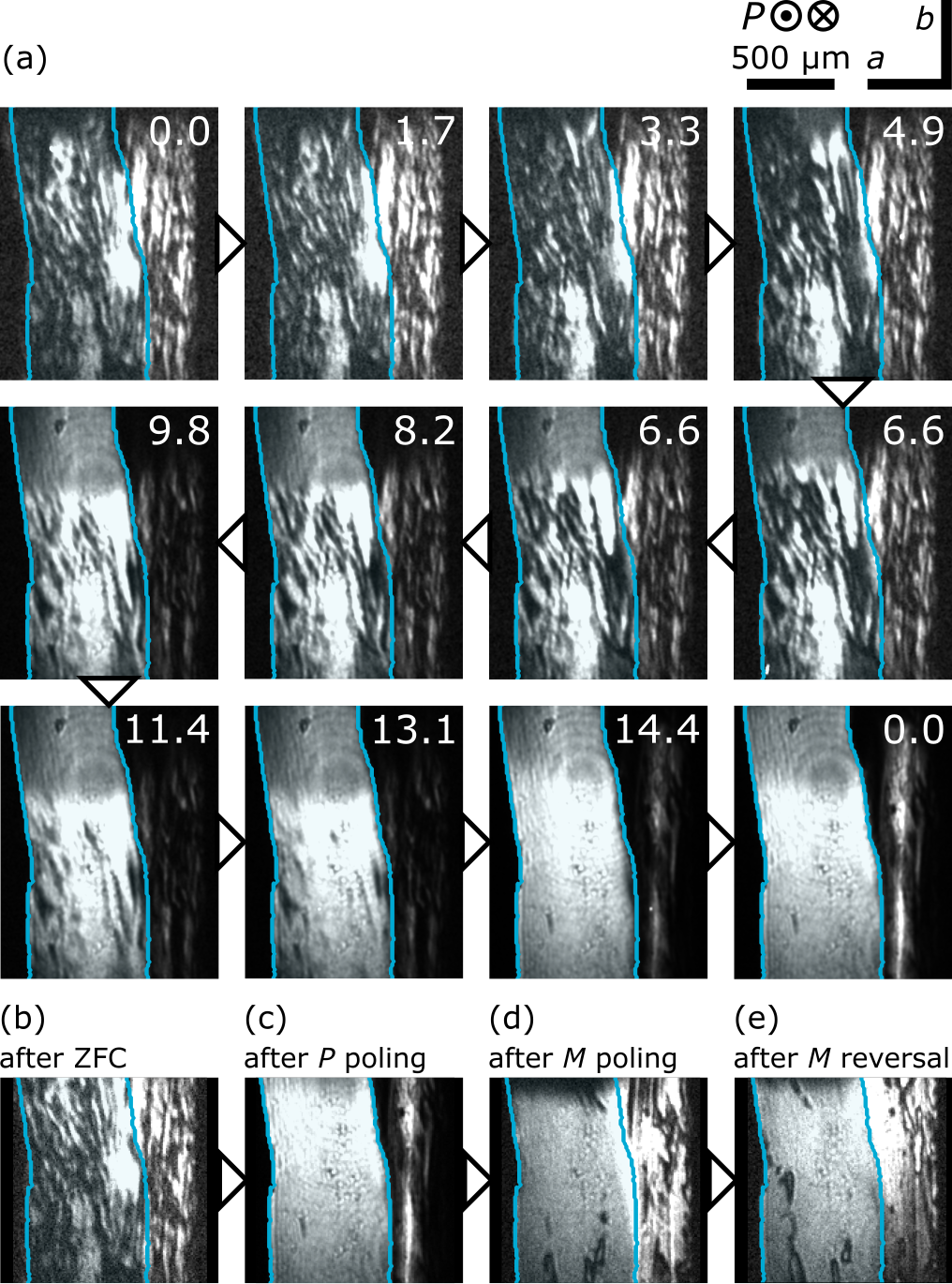}
	\caption{%
		\textbf{Electric-field-induced ferroelectric domain evolution.}
		(a-c)~Sequential images show the ferroelectric domain poling, starting from a multi-$P$--multi-$M$ domain pattern obtained after zero-field cooling with increasing electric field (upper right corner, in \si{MV/m}). In the region of overlap between front and back ITO electrodes (blue outline), a single-$P$--multi-$M$ domain pattern, indicated by its uniform brightness, is obtained after application of $E>\SI{14}{MV/m}$ and retained when the field is removed (last panel).
		(c-e)~The single-$P$--multi-$M$ domain pattern in (c), obtained after electric field poling, can be converted to a multi-$P$--single-$M$ domain pattern via the trilinear magnetoelectric effect in Eq.~(\ref{eq:trilinear}), as demonstrated by the reappearance of ferroelectric domains upon application of a magnetic field $H_c$ in (d) and (e).
	}
	\label{fig:electric-field-switching}
\end{figure}

The ferroelectric domains of \MnGeO\ can also be directly controlled via a conjgated applied electric field [see $F_\mathrm{E}$ in Eq.~(\ref{eq:bilinear})].
After zero-field cooling, the $c$-cut sample obtains a multi-$M$--multi-$P$ domain pattern, Fig.~\ref{fig:electric-field-switching}(a,b). 
Here, domains with opposite polarity $\pm P$ show up as white stripes with a typical width of \SIrange{20}{50}{\micro m} preferably aligned along the crystallographic $b$ direction, and separated by black domain walls as explained in Sec.~\ref{sec:optics}. 

Application of a sufficiently large electric field leads to the growth of a region with uniform brightness associated with a single polarization domain, as shown in Fig.~\ref{fig:electric-field-switching}(a). 
Approximately $\SI{5}{MV/m}$ are required to initiate domain growth, giving a lower bound for the coercive field to orient the ferroelectric domains.
The domain evolution is limited to the area in which the front and back ITO overlap (blue outline), whereas the material surrounding it remains unaffected.

Saturation is achieved for fields of about \SI{14}{MV/m}, and the resulting single-$P$ domain configuration is retained after the applied electric field is removed; see Fig.~\ref{fig:electric-field-switching}(c). 
Thus, application of an electric field converts a multi-$P$--multi-$M$ domain pattern, shown in Fig.~\ref{fig:electric-field-switching}(b), to a single-$P$--multi-$M$ domain pattern, shown in Fig.~\ref{fig:electric-field-switching}(c).

In addition to acting on the polarization $P$, due to the magnetoelectric cross-coupling term in Eq.~(\ref{eq:trilinear}), the electric field also acts on the magnetization domains of \MnGeO. 
This is revealed by the image sequence shown in Fig.~\ref{fig:electric-field-switching}(c-e): Starting from the electric-field-poled single-$P$--multiple-$M$ state, subsequent magnetic-field poling and magnetization reversal lead to a multi-$P$--single-$M$ domain pattern, resulting in the reappearance of domains indicated by black domain walls.


Assuming the magnetoelectric coupling term in Eq.~(\ref{eq:trilinear}) has been globally minimized after zero-field cooling, the final ferroelectric domain pattern after electric- and magnetic-field poling should correspond to the product of the initially largely independent $P$ and $M$ domain distributions. 
In this case, the observed domain size should be similar to the smaller size of the initial $P$ or $M$ domains. The prediction of smaller domains contradicts our experimental observations, however, as after the initializing magnetic-field cycle [Fig.~\ref{fig:electric-field-switching}(d,e)] the ferroelectric domains are larger than after zero-field cooling [Fig.~\ref{fig:electric-field-switching}(b)].

The observation of apparent domain-size evolution further corroborates our earlier conclusion that zero-field cooling into the multiferroic state of \MnGeO\ results in a metastable domain configuration. In this initial state, the order parameters $P$, $M$, and $\mathcal{C}$ do not yet conform globally to the energy minimization dictated by the trilinear magnetoelectric coupling term in Eq.~(\ref{eq:trilinear}). 
The first field cycle then leads to a minimization of the trilinear coupling term $F_\mathrm{ME}$ according to Eq.~(\ref{eq:trilinear}), accompanied by the formation of larger domains.
In addition, appearance of predominantly straight domain walls reveals a tendency to minimize the total length of the energy-wise costly domain boundaries.

\section{Discussion}
\label{sec:model}

\begin{figure}[tb] 
	\centering
	\includegraphics[width=86mm]{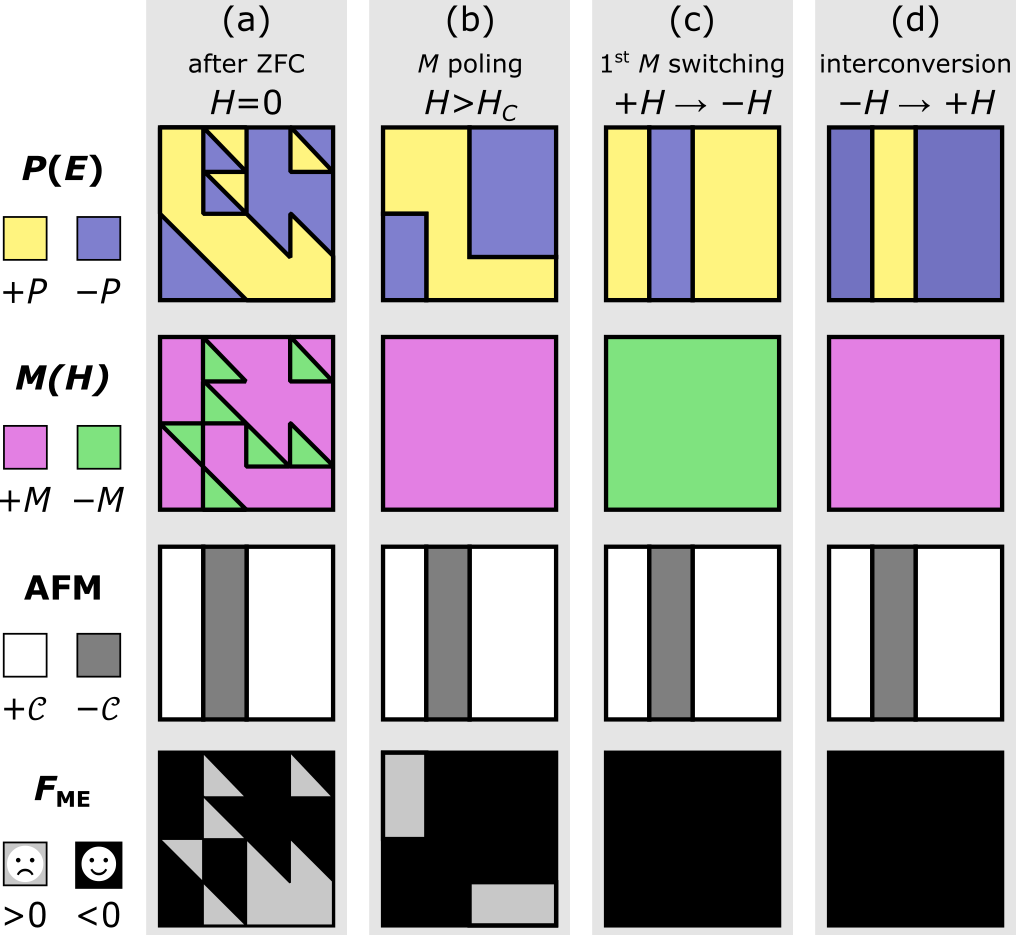}
	\caption{%
		\textbf{Phenomeonological model for magnetoelectric initalisation.}
		Pictograms show the spatial distribution of the observed polarization $P$ (top row, yellow/blue), of the accompanying magnetization $M$ (second row, pink/green), of the ``hidden'' antiferromagnetic domains characterized by $\mathcal{C}$ (third row, white/gray), as well as of the sign of the magnetoelectric coupling term $F_\mathrm{ME}$ in Eq.~(\ref{eq:trilinear}) (bottom row, gray/black).
		(a)~After zero-field cooling a metastable multi-$P$--multi-$M$ domain pattern is obtained, which does not yet minimize the trilinear magnetoelectric coupling term $F_\mathrm{ME}$. 
		(b-d)~A magnetoelectric initialisation procedure, combining poling of the intial multi-$M$ domain pattern and subsequent switching of the single-$M$ domain pattern, leads to a step-wise minimization of $F_\mathrm{ME}$ until an equilibrium domain configuration with global $F_\mathrm{ME}<0$ is obtained.
		%
		%
		Panels (a-d) [(b-d)]~corresponds to the zero-field cooling results shown in Fig.~\ref{fig:contrast-reversal}(a-d) [Fig.~\ref{fig:contrast-reversal}(f-h)].
		}
	\label{fig:ME_effect}
\end{figure}

\begin{figure*}[tb] 
	\centering
	\includegraphics[width=176mm]{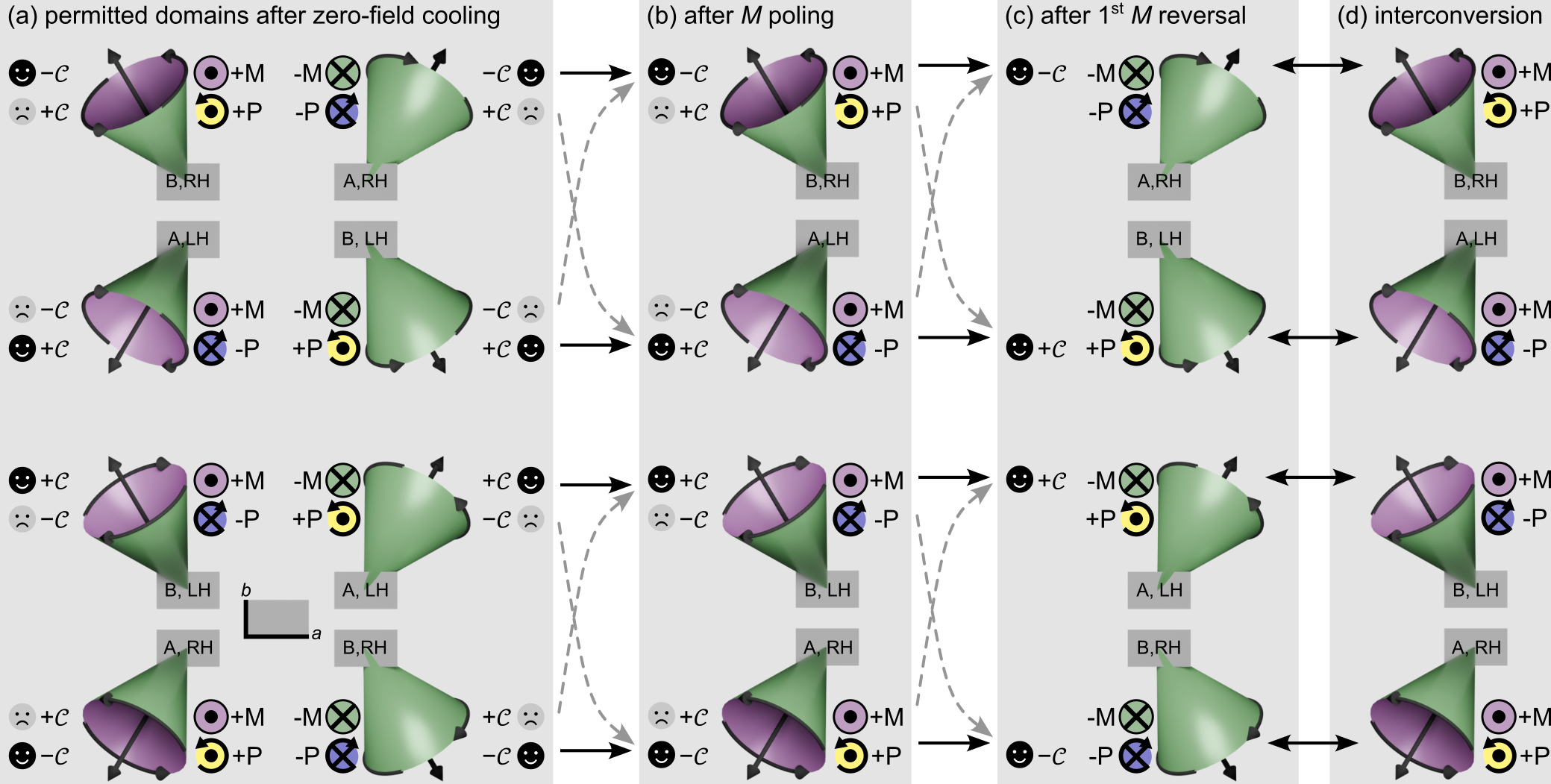}
	\caption{%
		\textbf{Microscopic model for deterministic domain initialization in \MnGeO.}
		Here, the spin-spiral state is represented by a single cone denoting the spiral component of one of the several conical spin spirals within the unit cell, adapted from \cite{2017Honda}. 
		The cone-axis orientation (central straight black arrow) determines the sign of the spontaneous magnetization $M$ (green/pink). The circulation around the cone with clockwise or counter-clockwise projections onto the $c$-plane, determines the sign of the spontaneous polarization $P$ (black arrows, green/blue). A and B denote translation domains associated with $\vec{Q}_{\rm A}$ and $\vec{Q}_{\rm B}$, respectively, and the sign of $\mathcal{C}$ determines the spiral chirality (left-handed: LH, right-handed: RH).
		(a)~After zero-field cooling, 16 possible domain states characterized by a combination of $P$, $M$ and $\mathcal{C}$ can arise, albeit only half of these minimize the trilinear magnetoelectric coupling term $F_\mathrm{ME}$ in Eq.~(\ref{eq:trilinear}). 
		(b,c)~A magnetoelectric two-step initialization procedure leads to a step-wise elimination of energetically unfavourable domain states (with $F_\mathrm{ME}>0$,{\large\frownie{}} ) from the sample via irreversible transformations (dashed gray arrows).
		(d)~Afterwards, an equilibrium single-$M$-multi-$P$ domain configuration with global $F_\mathrm{ME}<0$ ({\large\smiley{}}) is obtained, and the ferroelectric domain pattern can be repeatedly inverted upon magnetic-field reversal. 
	}
	\label{fig:ME_effect_microscopic}
\end{figure*}

In the following, we derive a phenomenological and a microscopic model to explain the irreversible, but deterministic, changes upon the magnetolectric initialization procedure.
We also describe key differences to domain ``annealing" or ``training" procedures commonly discussed in literature. 

At the heart of these models are two key observations:
First, ferroelectric polarization $P$ and ferromagnetic magnetization $M$ form different multi-domain patterns upon entering the multiferroic phase. 
Second, upon zero-field cooling, or cooling with only moderate fields applied, the system adopts a metastable configuration where the trilinear magnetoelectric coupling term $F_\textrm{ME}$ in Eq.~(\ref{eq:trilinear}) is not yet fully minimized.

The initial formation of largely independent domains in $P$, $M$, and $\mathcal{C}$ is facilitated by the first-order multiferroic transition in \MnGeO, which leads to the discontinuous, spontaneous emergence of the ordered state and their domains below $T_{\rm N}$. Consequently, higher-order terms in the free energy, such as the $F_\textrm{ME}$ in Eq.~(\ref{eq:trilinear}), are unlikely to be mimimized on a global scale when cooling in zero (or moderate) fields. 
%

The formation of a \textit{metastable} \textit{non-equilibrium} multiferroic domain pattern then requires a deterministic two-step field-poling protocol with subsequent global field switching to enable global and highly repeatable magnetoelectric domain inversion \cite{2012White,2017Honda,2018Leo_trilinear}.

\subsection{Generalised initialisation procedure}

On a phenomenological level, the effect of the magnetoelectric initialization procedure on the mesoscale domain distributions are illustrated in Fig.~\ref{fig:ME_effect}. Here, pictograms show the domain distributions of polarization $P$ (yellow/blue), magnetization $M$ (pink/green), and antiferromagnetic order $\mathcal{C}$ (white/gray), as well as the local sign of the trilinear magnetoelectric coupling term $\sigma(F_\mathrm{ME})=\sigma(P M \mathcal{C})$ (gray/black).
%
In regions where $\sigma(F_\mathrm{ME})>0$, highlighted in gray, the system is \textit{not} in a local equilibrium with respect to the magnetoelectric coupling term.

In the first step of the initialization protocol, Fig.~\ref{fig:ME_effect}(b), application of a magnetic field poles the multi-$P$--multi-$M$ domain pattern to a multi-$P$--single-$M$ state. At the same time, the system strives to minimize $F_\mathrm{ME}$, which can happen in two ways: First, if $\sigma(F_{\rm ME})<0$ is already in place, both $M$ and $P$ will switch simultaneously. Second, if $\sigma(F_\mathrm{ME})>0$ only the magnetic-field-driven order parameter $M$ will switch so that $\sigma(F_\mathrm{ME})<0$. As a result, the pattern of the $P$ domain observed after poling is \textit{not} a simple product between the initial $M$ and $P$ domains, and the change in domain morphology (that is, shape and size) is apparent, as also observed experimentally.

In the second step of the initialization protocol, Fig.~\ref{fig:ME_effect}(c), the magnetization of all the regions that have not reoriented under the first magnetic poling field are reversed. This gives the associated $P$ domains the opportunity to reverse -- or not -- toward the state of minimum energy, just as in the first step. Hence, by globally switching the magnetization from a single-$(+M)$ to a single-$(-M)$ state, the system can adopt a configuration that globally minimizes $F_\mathrm{ME}$, imprinting the domain pattern of antiferromagnetic $\mathcal{C}$ order onto the $P$ order.

If all possible domain states characterized by the combination of $P$, $M$ and $\mathcal{C}$ form with equal probability upon entering the multiferroic phase, a specific sequence of the switched $P$ domain fraction can be predicted: 
The first poling reverses 50\% of the $\pm M$ domains, and for 50\% of these, $\pm\rightarrow\mp P$ follows, which implies switching of only 25\% of $P$ domains. The subsequent field switch $+H\rightarrow -H$ reverses all $P$ domains from the first poling and half of the remaining $\pm P$ domains, leading to a switched domain fraction of 75\%. This step completes the initialization procedure, and henceforth, all $P$ domains (100\%) switch repeatedly with every magnetic-field reversal, as shown in Fig.~\ref{fig:contrast-reversal-supplement}. Although Fig.~\ref{fig:contrast-reversal}(a-d) implies such a trend, a quantitative analysis is prohibited by the small sample size in relation to the extension of the domains and the high density of crystal defects.

After this initialisation spanning exactly a full field cycle, the system has adopted a remanent domain configuration that leads to a global minimization of $F_\mathrm{ME}$ (highlighted in black). Subsequent switching of the magnetization, Fig.~\ref{fig:ME_effect}(d), then leads to repeatable magnetoelectric inversion of the ferroelectric multi-domain pattern in agreement with previously published results \cite{2012Honda,2017Honda,2018Leo_trilinear}.

\subsection{Magnetoelectric initialisation in \MnGeO}


The effects underlying the evolution of multiferroic domains in \MnGeO\ have not been studied earlier because integrated pyroelectric current and neutron scattering experiments required single-$P$--single-$M$ domain patterns to lead to meaningful results \cite{2012Honda,2012White,2017Honda}.
Note that the combined cooling fields applied in experiments of this type \cite{2017Honda}, that is, $H=\SI{1.5}{T}$ and $E=\SI{3}{MV/m}$ (and therefore $EH=\SI{4.5e6}{VT/m}$), are at least one order of magnitude larger than the moderate fields used for the SHG domain imaging experiments in this work, that is, $H^\text{max}=\SI{0.5}{T}$ and $E^\text{max}=\SI{0.6}{MV/m}$ (respectively, $(EH)=\SI{0.5e6}{VT/m}$). 

An interesting observation that might point to the influence of the non-equilibrium domain configuration on macroscopic measurements even after field cooling has been presented in \cite{2020Fischer}, where the authors report an irreversible polarization increase by 15\% upon the first application of a field.

On a microscopic level, the initialisation procedure is related to the magnetic-field-driven response of the complex conical spin-spiral structure of \MnGeO. The corresponding steps are schematically depicted in Fig.~\ref{fig:ME_effect_microscopic}. Here, a representative cone of the conical Mn spin spiral is shown. The orientation of the axis determines the sign of the magnetization $M$ (green/pink), and the projected circulation around the cone determining the sign of the polarization $P$ (black arrows, green/blue). 
In ground-state equilibrium, each incommensurate domain state characterized by $\vec{Q}_{\rm A}$ or $\vec{Q}_{\rm B}$ (just ``A'' and ``B'' in Fig.~\ref{fig:ME_effect_microscopic}) can exist in a left- or right-handed chirality, determined by the sign of $\mathcal{C}$ \cite{2017Honda} [Fig.~\ref{fig:ME_effect_microscopic}(c,d)].

After zero-field cooling across the first-order transition, domains of any of the  combinations of $M$, $P$, and incommensurate translation vectors, A or B, can form. 
Furthermore, there are two cone orientations, respectively $\mathcal{C}$, for each of these sign combinations, resulting in a total of 16 possible initial domain states. Eight of these are schematically represented in Fig.~\ref{fig:ME_effect_microscopic}(a). The choice of $+\mathcal{C}$ or $-\mathcal{C}$ for each of these is indicated by icons to the left or right.

If the relative sign of $M\cdot P$ is opposite to the sign of $\mathcal{C}$ (indicated with {\large\smiley{}}), we have a ground-state configuration, where Eq.~(\ref{eq:trilinear}) is minimized. A switching field then leads to a \textit{reversible} rotation of the cone axis, inverting both $P$ and $M$ simultaneously, as well as interchanging the $\vec{Q}_{\rm A}$ and $\vec{Q}_{\rm B}$ domains, while leaving the chirality (and therefore $\mathcal{C}$) unchanged (solid black arrows). 

In contrast, if $\sigma(P\cdot M)=\sigma(\mathcal{C})$, the domain state is energetically unfavorable (indicated with {\large\frownie{}}), and a likely \textit{irreverisble} transformation under applied fields is a reorientation of the cone axis via a ``cone flop'' (similar to an umbrella flipping over). This reverses $M$, but leaves the translation vector, $\vec{Q}_{\rm A}$ or $\vec{Q}_{\rm B}$, as well as the polarity (i.e., projected circulation) unchanged. As this operation changes the sign of $M$ only, the trilinear energy term is then minimized.

\subsection{Differences to domain ``training" cycles}

The fully deterministic step-wise initialisation procedure described in this work differs fundamentally from so-called domain ``training" procedures commonly associated with improving the hysteretic response in ferrolectric materials \cite{2013Zhou, 2018DeLuca, 2021Muller, 2023Saini}:
For these, field or temperature cycling redistributes domain walls to energetically favourable positions and removes trapped charges at the electrodes. These \textit{stochastic} processes gradually change the electrochemical sample properties dependent on \textit{local} energy contributions and require \textit{repeated poling cycles} to approach a stable response.

In contrast, the procedure shown here for \MnGeO\ is a fully \textit{deterministic} process that requires a \textit{single} full (half) field cycle to a fully reversible remanent domain state when starting from a zero-field-cooled (single-field-cooled) initial domain state: 
The irreversible changes in the domain distribution are driven by the step-wise field-driven \textit{global} minimisation of the trilinear magnetoelectric coupling term. The irreversible switching processes involved explain the change in the observed domain morphology, i.e., preferred domain boundary orientations and apparent domain sizes, and the expected switching fractions of 25\%$\rightarrow$75\%$\rightarrow$100\% in each step.

The magnetoelectric intialisation procedure demonstrated here for the example of \MnGeO\ is key to obtain a domain configuration that shows the repeated domain interconversion mediated by the trilinear coupling term of Eq.~(\ref{eq:trilinear}). 
Such strong and reversible coupling between spontaneous order parameters allows for deterministic magnetoelectric cross-control and is robust against disruptive effects such as strain or magnetic stray fields.

We expect that similar initialisation procedures are relevant in other materials featuring a complex configuration of order parameters, and where a first-order phase transition or rapid quenching through a second-order phase transition promote non-equilibrium domain configurations. 
Similar effects could, for example, occur in bulk multiferroics such as spinel CoCr$_2$O$_4$ \cite{2006Yamasaki}, orthoferrites \cite{2009Tokunaga,2021Hassanpour}, or hexaferrites \cite{2012Chun,2019Chmiel,2019Ueda}, as well as artificial multiferroic heterostructures and interfaces with engineered trilinear coupling terms \cite{2018DeLuca,2019Strkalj}. 
The presented understanding of the intialisation field protocol and switching processes to achieve reliable magnetoelectric cross-control are thus key to harnessing these materials for future applications.

\section{Conclusions}
\label{sec:conclusions}

In this work we used optical SHG imaging to show how the repeatable domain interconversion of multiferroic domains in \MnGeO\ can only be achieved by performing a deterministic initialization procedure. 
This procedure spans a single full field cycle after zero-field-cooling, with irreversible domain transformations being driven by the step-wise minimisation of the global magnetoelectric coupling term. This fully deterministic behaviour in \MnGeO\ is in marked contrast to often-discussed ferroelectric ``training" which relies on the stochastic motion of domains under repeated field or temperature cycles.

The understanding of the domain initialisation protocol is key to future reliable magnetoelectric functionalities based on multiferroic bulk and composite materials.


\begin{acknowledgments}
	
	The authors thank Morgan Trassin for sputtering the ITO electrodes and Roman V. Pisarev for helpful discussions.
	JSW and MK acknowledge funding from the Swiss National Science Foundation (SNSF) through grants number 200021\_153451 and 200021\_165855.
	TH and TK acknowledge funding from JSPS KAKENHI Grant Numbers JP17H01143 and JP21H04436.
    %
    DM acknowledges funding from the Onsager Fellowship Program and the Outstanding Academic Fellow Program. His work was in part supported by the Research Council of Norway through its Center of Excellence funding scheme, project number 262633, ``QuSpin''.
    MF acknowledges funding by the SNSF through grants number 200021\_147080 and 200021\_178825 and appreciates support by ETH Zurich.
	%
	NL was supported by a UKRI Future Leaders Fellowship [grant number MR/X033910/1].
	
\end{acknowledgments}



%

\end{document}